\documentclass[preprint,showpacs,prd]{revtex4-1}
\usepackage{epsfig,amsmath}
\usepackage{graphicx}
\usepackage{mathrsfs}
\usepackage{bm}

\begin{document}
\title{\textbf{The interference effect of laser-assisted bremsstrahlung emission in Coulumb fields of two nuclei}}

\author{Ankang Li$^{1}$, Jiaxiang Wang $^{1,a}$, Na Ren$^{1}$, Pingxiao Wang$^{2}$,Wenjun Zhu${^3}$, Xiaoya Li${^3}$, Ross Hoehn$^{4}$, S. Kais$^{4,5}$}

\affiliation{$^{1}$State Key Laboratory of Precision Spectroscopy and Department of Phyasics, East China Normal University, Shanghai 200062, China}

\affiliation{$^{2}$Applied Ion Beam Physics Laboratory, Key Laboratory of the Ministry of Education ,
China and Institute of Modern Physics, Department of Nuclear Science and Technology,
Fudan University, Shanghai 200433, China}

\affiliation{$^{3}$National Key Laboratory of Shock Wave and Detonation Physics, Mianyang 621900, Sichuan, China}

\affiliation{$^{4}$Departments of Chemistry and Physics, Purdue University, West Lafayette, Indiana 47907, USA}
\affiliation{$^{5}$Qatar Environment and Energy Research Institute, Qatar Foundation, Doha, Qatar}

\email{$^{a)}$ Author to whom correspondence should be addressed. Electronic mail:jxwang@phy.ecnu.edu.cn}

\def\abstractname{}
\begin{abstract}

In this paper, the spontaneous bremsstrahlung emission from an
electron scattered by two fixed nuclei in an intense laser field is
investigated in details based upon the Volkov state and the
Dirac-Volkov propagator. It has been found that the fundamental
harmonic spectrum from the electron radiation exhibits distinctive
fringes, which is dependent not only upon the internucleus distance
and orientation, but also upon the initial energy of the electron
and the laser intensity. By analyzing the differential cross
section, we are able to explain these effects in terms of
interference among the electron scattering by the nuclei. These
results could have promising applications in probing the atomic or
molecular dressed potentials in intense laser fields.

\end{abstract}
\maketitle
\section*{\textbf{\romannumeral1. Introduction}}

High-order harmonic generation (HHG) \cite{1,2,3,4} is a process in
which high-order harmonics of the fundamental laser frequency are
coherently radiated when an intense laser pulse is focused into an
atomic or molecular gas. This process is not only used to generate
UV or XUV lights, but also be applied to explore molecular
structures, recently. The first breakthrough was the discovery of a
double-slit-type interference effect from the simplest diatomic
molecules {${H_2^+}$} and $H_2 $\cite{5,6,7,8}. The experimental
confirmation was first realized for aligned {$CO_2$} in 2005
\cite{9,10}. The next major development was the so-called the
molecular orbital tomography proposed by Itatani et in 2004
\cite{11}. Namely, once the HHG spectra and phases are known for
various orientation of molecular axis, a 2D projection of the
initial electron orbital can be reconstructed through a tomographic
algorithm. Now, the work has been generalized to include orbital
symmetry influences upon HHG and quantum tomography with 2D
calculations\cite{12}.

In all the above-mentioned work, the HHG originates from the
electrons bound by the atoms or molecules and the calculation
usually involves the time-dependent Schrodinger equation (TDSE) with
dipole approximation. But when the field is so strong that the
ponderomotive energy of the free electron reaches the same order of
the rest energy of the electron, there will be a different picture.
Namely, the dipole approximation may not a good choice and the TDSE
should be replaced by the Dirac equation. Moreover, some electrons
may be ionized to be free particles, whose dynamics will be
predominated by the intense laser fields instead of the Coulomb
potentials. Now, the principle process is the so-called
laser-assisted bremsstrahlung, which has been studied previously by
several authors. In the early works, the analytic expression for the
radiation spectrum of laser-assisted bremsstrahlung in a plane
monochromatic has been derived by Karapetyan and Fedorov for
nonrelativistic regime \cite{13}. Within the framework of the Born
approximation, Roshchupkin \cite{14,15}has developed a general
relativistic expression for the amplitude of the scattering of an
electron by a nucleus in an external field with arbitrary intensity.
Recently, the numerical evaluation of the laser-assisted
bremsstrahlung process has been carried out for both circularly
polarized and linearly polarized laser field\cite{16,17}.

Motivated by the molecule HHG in non-relativistic case, in this
paper we will consider an electron scattering by two nuclei in
strong laser fields. This model differs from one-nucleus case
mentioned above by providing more than one center for the electron
scattering, which will allow for dynamics, for example, the emission
spectra of the electron may depend on the internuclear distance and
orientation, just as in the situation of molecule HHG. This model
could provide us a method to explore some special potentials, which
exists only in intense laser fields, such as the dressed
Kramer-Henneberg potential in high-frequency laser fields, which
plays an important role in guaranteeing the existence of
multiply-charged negative ions in the fields.

The notations used in this paper are as follows. The four-vector
product is denoted by {$a\cdot b=a^0b^0-\bm{ab}$}. For the Feynman
dagger, we use the following notation: {$\rlap{\slash}A=\gamma\cdot
A$}. The Dirac adjoint is denoted by the standard notation
{$\overline{u}=u^{\dagger} \gamma^0$} for a bispinor u and
{$\overline{F}=\gamma^0 F^{\dagger}\gamma^0$} for a matrix F.

The outline of this paper is the following. First, we will introduce
the laser-assisted bremsstrahlung model and derive the theoretical
expression for the cross section of the emission in Sec
\romannumeral2. Then, the numerical estimation of the cross section
and the corresponding analyses will be provided in Sec
\romannumeral3. Concluding remarks are reserved for Sec
\romannumeral4.

\section*{\romannumeral2. Theoretical derivation of the Bremsstrahlung cross section}

Consider two nuclei with charge number Z are fixed in the x-z plane
with an internucleus distance $R_0$ in a strong laser field. We
assume that, in the laboratory frame of reference, the laser can be
described by a plane wave propagating in the positive direction of
the z-axis with a vector potential {$A^{\mu}$}:

\begin{eqnarray}
{A^{\mu}}&=&A_0[{\delta}\cos\phi{\epsilon_1}^\mu+(1-\delta^2)^{1/2}\sin\phi{\epsilon_2}^\mu],\label{field}
\label{plane}
\end{eqnarray}

The approximation is acceptable if the number of laser photons are
large enough so that an arbitrary amount of energy and momentum can
be taken from or emitted into the field without changing it. The
plane wave depends only on the the phase factor {$\phi=k\cdot x$},
in which $x$ is the position vector, and
{$k^\mu=\frac{\omega_0}{c}(1,0,0,1)$} the four wave vector with
$\omega_0$ denoting the laser frequency. The laser is circularly
polarized for $\delta={1}/{\sqrt{2}}$ and linearly polarized for
$\delta=0,\pm1$. We define two polarization vectors $\epsilon_1$,
$\epsilon_2$, satisfying $\epsilon_i\cdot k=0,
\epsilon_i\cdot\epsilon_j =\delta_{ij}$ $(i,j=1,2)$. The laser
intensity can be easily described by a dimensionless parameter
$Q=eA_0/(mc^2)$, which is usually called laser intensity parameter.
It should be mentioned that in the nonrelativistic regime, the
characteristic velocity and energy for an electron moving in such an
electromagnetic field is $v\sim{eA_0/(mc)}$ and
$E\sim{e^2A_0^2/(mc^2)}$, so a relativistic treatment is necessary
if $v\sim c$ and $E\sim{mc^2}$ is satisfied, which means the motion
of the electron will become relativistic when $Q\sim1$ .

The angle between the orientation of two nuclei and the laser
propagation direction is denoted by $\vartheta$. For convenience of
calculation, here we set the origin of the coordinate at the middle
of two nuclei. So we can easily introduce a vector
$\bm{R}=R_0(\sin\vartheta\bm{e_x}+\cos\vartheta\bm{e_z})/2$ to
describe the location of the two nuclei.

\begin{figure}
\includegraphics[scale=0.5]{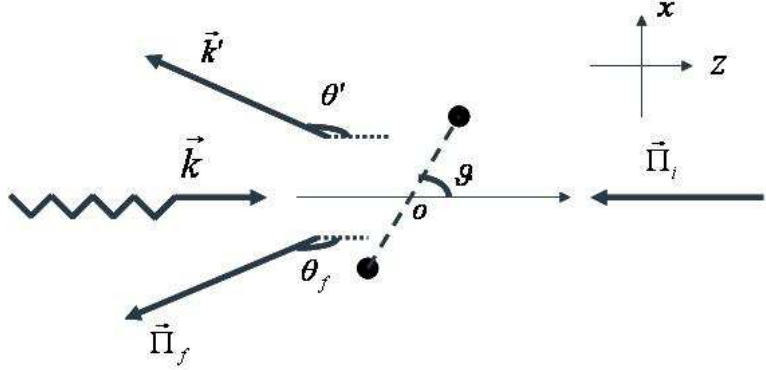}
\caption{\label{fig:geometry}The scattering geometry: The incoming
electron with laser-dressed four momentum $\Pi_i$ counterpropagates
with the laser while scattering by two fixed nuclei. $\vartheta$ is
an angle between internucleus axis and the laser propagating
direction. The final electron with $\Pi_f$ and the bremsstrahlung
photon with $k^\prime$ are projected onto the xz plane in this
figure; So only the polar angles $\theta_f$ and $\theta^\prime$ are
displayed. The azimuthal angles are denoted by $\Omega_f$ and
$\Omega^\prime$, respectively.}
\end{figure}

\begin{figure}
\includegraphics[scale=0.5]{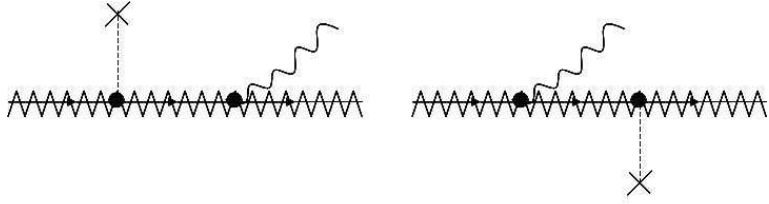}
\caption{\label{fig:Feynman} Feynman diagrams describing
laser-assisted bremsstrahlung. The laser-dressed electron and
laser-dressed electron propagator are denoted by a zigzag line on
top of the straight line. The Coulomb field photon is drawn as a
dashed line, and the bremsstrahlung photon as a wavy line.}
\end{figure}

Now we begin to derive the differential cross section of the
electron-nucleus bremsstrahlung. Consider the scattering geometry
that an incoming electron moving along the negtive z-axis has a
head-on collision of the laser photons while scattering by two
nuclei. The configuration is shown in Fig. ~\ref{fig:geometry}. The
whole process can be described by two Feynman diagrams displayed in
Fig. ~\ref{fig:Feynman}. In the first one, the initial electron
first interacts with two nuclei and then emits a bremsstrahlung
photon. The situation is reversed in the second diagram. In Feynman
diagrams, the electron is denoted by a zigzag line on top of a
straight line since it is dressed by a strong laser. Also here the
free electron propagator is replaced by the Dirac-Volkov propagator
\cite{18}.
%

Actually, the electron will interact with three external fields
during the process, namely, the laser field described by (1), the
Coulomb field of two nuclei and the field of the emitted
bremsstrahlung photon. As usual, we treat the laser-electron
interaction exactly and nonperturbatively by using Volkov states as
the initial and final wave functions:

\begin{eqnarray}
\psi_{p,r}&=&\sqrt{\frac{mc}{\Pi^0V}}\zeta_p(x)u_r(p), \\
\zeta_p(x)&=&\big(1+\frac{e\rlap{$\slash$}k\rlap{$\slash$}A}{2p\cdot
k}\big)e^{iS},
\end{eqnarray}

\begin{eqnarray}
S=&-&\frac{\Pi\cdot x}{\hbar}-\frac{e^2A_0^2}{8\hbar
c^2(p\cdot k)}(2\delta^2-1)\sin2\phi+\frac{eA_0}{\hbar c(p\cdot k)}\nonumber\\
&\times&\big[\delta({p}\cdot{\epsilon_1})sin\phi
-(1-\delta^2)^{1/2}({p}\cdot{\epsilon_2})\cos\phi\big].
\end{eqnarray}

Here $p$ is the four-momentum of the electron outside the field, and
{$\Pi=p+\frac{e^2A_0^2}{4c^2(p\cdot k)}$} is the corresponding
laser-dressed four-momentum, $u_r(p)$ the free Dirac spinor. Here we
employ a box normalization with a normalized volume V.

The interaction with the emitted radiation and Coulomb field is
taken to the first perturbation, in which the interaction between
electron and nuclei is considered under Born approximation:
$v_i/c\ll\alpha Z$. Here $\alpha$ is fine structure constant and
$v_i$ is the initial velocity of the electron. As to the Coulomb
field of the nuclei, we use a Yukawa potential with a screen length
$l_0$ instead of the conventional Coulomb potential to avoid
possible singularity at resonance. The four-vector potential of the
two fixed nuclei can be written as:

\begin{eqnarray}
A_Y^\mu&(\bm{r})=&-\frac{Ze\delta^{\mu0}}{|\bm{r}-\bm{R}|}e^{{|\bm{r}-\bm{R}|}/{l_0}}-\frac{Ze\delta^{\mu0}}{|\bm{r}+\bm{R}|}e^{{|\bm{r}+\bm{R}|}/{l_0}}.
\end{eqnarray}

The corresponding Fourier transform is:
\begin{eqnarray}
A_Y^\mu&(\bm{q})=-\frac{4\pi Ze}{\bm{q}^2+l_0^2}
(e^{i\bm{q}\bm{R}}+e^{-i\bm{q}\bm{R}}).
\end{eqnarray}
As we can see, the Fourier transform of the potential depends on the
inter-distance and the orientation of the two nuclei. This is the
origin of the interference effect on the radiation spectrum. The
four-vector potential of the emitted bremsstrahlung photon has the
form:

\begin{eqnarray}
A_c^\mu&(x)=\sqrt{{2\pi\hbar}/{\omega^\prime}}c{\epsilon_c}^\mu
e^{i{k^\prime} x}.
\end{eqnarray}

The wave vector of the emitted photon with polarization is described
by {$k^\prime=\frac{\omega^\prime}{c}(1,\bm{e_{k^\prime}})$},
$\bm{e_{k\prime}}=\cos\varphi^\prime \sin\theta^\prime
\bm{e_x}+\sin\varphi^\prime \sin\theta^\prime
\bm{e_y}+\cos\theta^\prime \bm{e_z}$. So the transition amplitude of
an electron scattering by two fixed nuclei in a strong laser field
can be specified by the following expression:
\begin{eqnarray}
S_{fi}=-\frac{e^2}{\hbar^2 c^2}{\int
dx^4dy^4\bar{\psi}_{p_f,r_f}(x)[\rlap{\slash}A_c(x)iG(x-y)\rlap{$\slash$}A_Y(y)
+\rlap{\slash}A_Y(x)iG(x-y)\rlap{$\slash$}A_c(y)]
\psi_{p_{i},r_{i}}(y)}\nonumber .\\
\end{eqnarray}

Here $iG(x-y)$ is the laser-dressed propagator of the electron,
which can be written as:

\begin{eqnarray}
iG(x-y)=-\int \frac{dp^4}{(2\pi\hbar)^3(2\pi
i)}\zeta_p(x)\,\frac{\rlap{\slash}p+mc}{p^2-m^2c^2}\,\bar{\zeta}_p(y).
\end{eqnarray}

Since we are not interested in investigating polarization or spin
properties, we average over the spin of the incoming electron, and
sum over the spin and polarization of the final electron. The
differential cross section is calculated with the formula:

\begin{eqnarray}
d\stackrel{\sim}{\sigma}=\frac{1}{2JT}\sum_{r_i,r_f,\varepsilon_c}
\big|S_{fi}\big|^2\frac{Vd^3\Pi_f}{(2\pi\hbar)^3}\frac{d^3k^\prime}{(2\pi)^3}.
\end{eqnarray}

Here T is the long observation time and
{$J=\frac{c}{V}\frac{\bm{\Pi_i}}{\Pi_i^0}$} stands for the incoming
particle flux. We have
$d^3\Pi_f=|\bm{\Pi_f}|^2d\Omega_f=|\bm{\Pi_f}|^2sin\theta_fd\theta_fd\varphi_f$,
$d^3k^{\prime}=\frac{{\omega^{\prime}}^2}{c^2}d\Omega^{\prime}=\frac{{\omega^{\prime}}^2}{c^2}sin\theta^{\prime}d\theta^{\prime}d\varphi^{\prime}$,\,
where $\Omega^{\prime}$ and $\Omega_f$ are solid angle for the
emitted photon and electron, respectively. Finally, we can derive
the expression of the average differential cross section for
emission or absorption of n photons as (for details, see Appendix A):
\begin{eqnarray}
&&\frac{d\stackrel{\sim}{\sigma}}{d\omega^\prime d\Omega^\prime
d\Omega_f}=\frac{\alpha(Z\alpha)^2}{8\pi^2c^2}\sum_{n,\varepsilon_c}
\frac{|\bm{\Pi_f}|}{|\bm{\Pi_i}|}\big|e^{i\bm{q}\bm{R}}+e^{-i\bm{q}\bm{R}}\big|^2\nonumber\\
&&\times\frac{\omega^\prime}{(\bm{q}^2+l_0^2)^2}Tr[\bar{R}_{fi,n}(p_f+mc)R_{fi,n}(p_i+mc)],\nonumber\\
\end{eqnarray}

where:
\begin{eqnarray}
R_{fi,n}&=&\sum_s
M_{-n-s}(\rlap{$\slash$}\epsilon_c,\eta^1_{\Pi,\Pi_f},\eta^2_{\Pi,\Pi_f})
\frac{i}{\rlap{$\slash$}p-mc}\nonumber\\
&&\times\bar{M}_{-s}(\gamma^0,\eta^1_{\Pi,\Pi_i},\eta^2_{\Pi,\Pi_i})\nonumber\\
&&+\sum_{s^\prime}M_{-n-s^\prime}(\gamma^0,\eta^1_{\Pi^\prime,\Pi_f},\eta^2_{\Pi^\prime,\Pi_f})\frac{i}{\rlap{$\slash$}p^\prime-mc}\nonumber\\
&&\times\bar{M}_{-s^\prime}(\rlap{$\slash$}\epsilon_c,\eta^1_{\Pi^\prime,\Pi_i},\eta^2_{\Pi^\prime,\Pi_i}),\nonumber\\
\end{eqnarray}

with the argument defined as:
\begin{eqnarray}
\eta^1_{p_1,p_2}&=&\frac{eA_0}{\hbar
c}\delta[\frac{{p_2}\cdot{\epsilon_1}}{k\cdot
p_{2}}-\frac{{p_1}\cdot{\epsilon_1}}{k\cdot p_1}],\nonumber\\
\eta^2_{p_1,p_2}&=&-\frac{eA_0}{\hbar
c}(1-\delta^2)^{1/2}[\frac{{p_2}\cdot{\epsilon_1}}{k\cdot
p_{2}}-\frac{{p_1}\cdot{\epsilon_1}}{k\cdot p_1}],\nonumber\\
\end{eqnarray}

The four-momentum transfer onto the Coulomb field by two fixed
nuclei is denoted by $q^{\mu}=(0,\bm{q})$ , and the two
laser-dressed four-momenta of the virtual electrons in the Feynman
diagrams by $\Pi,\Pi^\prime$. They are given by the energy-momentum
conserving relation during the scatting process:

\begin{eqnarray}
\Pi&=&\pi_f-(n+s)\hbar k+\hbar k^\prime,\nonumber\\
\Pi^\prime&=&\pi_i-s\hbar k-\hbar k^\prime,\nonumber\\
\hbar q&=&\pi_f-\pi_i+\hbar k^\prime-n\hbar k,\nonumber\\
\end{eqnarray}

M is a $4\times4$ matrix with five arguments:
\begin{eqnarray}
&&M_{s}(F,\eta^1_{p_1,p_2},\eta^2_{p_1,p_2})=\nonumber\\
&&\big[\rlap{$\slash$}F+\frac{e^2{A_0}^2}{8c^2}\frac{\rlap{$\slash$}k\rlap{$\slash$}F\rlap{$\slash$}k}{(p_i\cdot
k)(p_2\cdot k)}\big]G^0_s(\alpha,\beta,\varphi)\nonumber\\
&&+\frac{eA_0}{2c}\delta\big[\frac{\rlap{$\slash$}\epsilon_1\rlap{$\slash$}k\rlap{$\slash$}F}{(p_1\cdot
k)}+\frac{\rlap{$\slash$}F\rlap{$\slash$}k\rlap{$\slash$}\epsilon_1}{(p_2\cdot
k)}\big]G^1_s(\alpha,\beta,\varphi)\nonumber\\
&&+\frac{eA_0}{2c}(1-\delta^2)^{1/2}\big[\frac{\rlap{$\slash$}\epsilon_2\rlap{$\slash$}k\rlap{$\slash$}F}{(p_1\cdot
k)}+\frac{\rlap{$\slash$}F\rlap{$\slash$}k\rlap{$\slash$}\epsilon_2}{(p_2\cdot
k)}\big]G^2_s(\alpha,\beta,\varphi)\nonumber\\
&&+(\delta^2-\frac{1}{2})\frac{e^2{A_0}^2}{4c^2}\frac{\rlap{$\slash$}k\rlap{$\slash$}F\rlap{$\slash$}k}{(p_i\cdot
k)(p_2\cdot k)}G^3_s(\alpha,\beta,\varphi),\nonumber\\
\end{eqnarray}

The generalized Bessel functions are given by:
\begin{eqnarray}
G^0_s(\alpha,\beta,\varphi)&=&\sum_n
{J_{2n-s}(\alpha)J_n(\beta)e^{i(s-2n)\varphi}},\nonumber\\
G^1_s(\alpha,\beta,\varphi)&=&\frac{1}{2}\big(G_{s+1}^0(\alpha,\beta,\varphi)+G_{s-1}^0(\alpha,\beta,\varphi)\big),\nonumber\\
G^2_s(\alpha,\beta,\varphi)&=&\frac{1}{2i}\big(G_{s+1}^0(\alpha,\beta,\varphi)-G_{s-1}^0(\alpha,\beta,\varphi)\big),\nonumber\\
G^3_s(\alpha,\beta,\varphi)&=&\frac{1}{2}\big(G_{s+2}^0(\alpha,\beta,\varphi)+G_{s-2}^0(\alpha,\beta,\varphi)\big).\nonumber\\
\end{eqnarray}

With the corresponding argument:
\begin{eqnarray}
\alpha&=&[(\eta^1_{p_1,p_2})^2+(\eta^2_{p_1,p_2})^2]^{1/2},\nonumber\\
\beta&=&\frac{Qm^2c^2}{8\hbar}(2\delta-1)\big(\frac{1}{k\cdot
p_1}-\frac{1}{k\cdot p_2}\big)\nonumber\\
\varphi&=& arctan(-\frac{\eta^2_{p_1,p_2}}{\eta^1_{p_1,p_2}}).\nonumber\\
\end{eqnarray}

The differential cross section in (11) is evaluated for both the
direction of the final electron and the bremsstrahlung photon. Here
we are more interested in the influence by the internuclear distance
and orientation on the bremsstrahlung photon spectrum, so we
integrate the differential cross section over the solid angle
$\Omega_f$ of the outgoing electron and obtain a cross section
differential only in the direction of the emitted bremsstrahlung
photon and its energy:

\begin{eqnarray}
\frac{d\sigma}{d\omega^\prime d\Omega^\prime}=\int
\frac{d\stackrel{\sim}{\sigma}}{d\omega^\prime d\Omega^\prime
d\Omega_f}d\Omega_f.
\end{eqnarray}

It is well known that the resonance occurs when the intermediate
electron fall within the mass shell\cite{14,15,16,17}. That's
because the lower order processes (here refers to the nonlinear
Compton scattering) is allowed in the field of a light wave.
Although the resonance is a characteristic feature of the
second-order process like bremsstrahlung, but it will not draw much
of our attention here since the cross section (18) at resonance will
not be affected by the internuclear distance or orientation, for
which the screening length need not be discussed here. More details
will be given in the next section.
%

\section*{\romannumeral3. Numerical Results}

\begin{figure}
\includegraphics[scale=0.5]{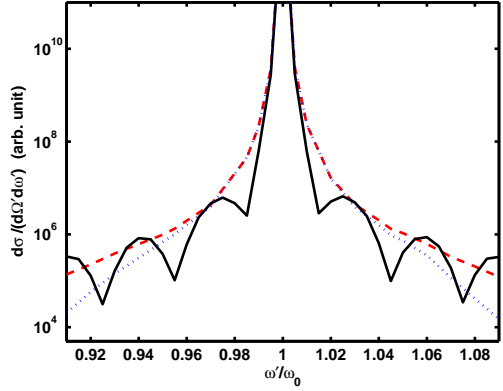}
\caption{\label{fig:cross-section} The cross section for the
fundamental harmonic at $\theta^\prime=1^\circ$. Here we consider an
electron with initial energy 5MeV head-collide with a circularly
polarized laser with intensity parameter $Q=17.8$ and is scattered
by two fixed nuclei. The internucleus distance is 1nm for the full
line, 0.152nm for the dotted line, 0.1nm for the dash line.}
\end{figure}

In this section, we will present some examples of the cross section
in (18) for different internuclear distance or orientation. We
consider the internuclear distance of two fixed proton $(Z=1)$ is
about several atom units. To observe the affect of the Coulomb field
of two fixed nuclei on the spectra, we have to choose the laser
frequency in an X-rays order: the wavelength is 0.2nm. The intensity
of the laser is $Q=17.8$ and circularly polarized. First we consider
the electron has an initial energy of $E_i=5MeV$ and the orientation
of the two nuclei is parallel to the direction of laser propagation.
The cross section of the fundamental harmonic for scattering angle
$\theta^\prime=1^\circ$ is shown in Fig ~\ref{fig:cross-section}.
The most remarkable feature of the spectrum is that there are minima
at some frequencies for large internuclear distance.

\begin{figure}
\includegraphics[scale=0.5]{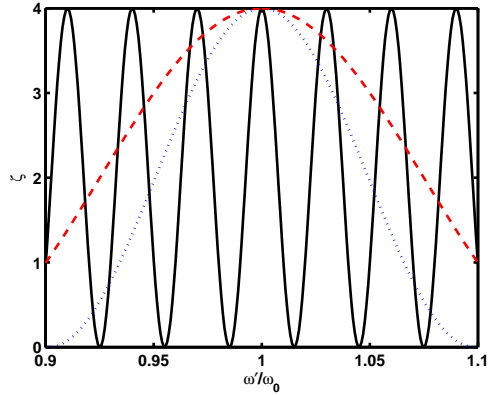}
\caption{\label{fig:mechanism}The relation between the parameter
$\zeta(\bm{q},\bm{R})$ and the bremsstrahlung photo frequency
$\omega^\prime$ at $\theta_f=\pi$  for absorbing 1 photon in the
whole process (n=1) . The parameter of the electron and laser is the
same with Fig~\ref{fig:cross-section}. The internucleus distance is
1nm for the full line, 0.152nm for the dotted line, 0.1nm for the
dash line.}
\end{figure}

%
The mechanism behind this phenomenon is two-centre interference
during the scattering process, which is described by the term
$\zeta(\bm{q},\bm{R})=\big|e^{i\bm{q}\bm{R}}+e^{-i\bm{q}\bm{R}}\big|^2\sim
\cos^2(\bm{q}\cdot\bm{R})$ in the cross section. So when the
momentum transfer from the Coulomb field is so large that
$\bm{q}\cdot\bm{R}\sim1$ , the differential cross section in (11)
will be suppressed for some special condition. Since most of the
contribution to the integrand (18) comes from a small cone in the
forward direction of the ingoing electron ($\theta_f=\pi$), the
positions of the minima found in the spectra are largely determined
by the parameter $\zeta(\bm{q},\bm{R})$  in the backscattering
direction. This can be confirmed in Fig ~\ref{fig:mechanism}, which
plots $\zeta(\bm{q},\bm{R})$ as a function of harmonic frequency for
emission angle $\theta_f=\pi$. The positions of the minima in the
spectra are almost coincident with those of $\zeta(\bm{q},\bm{R})$,
which can be expected for the condition:
\begin{eqnarray}
{R_0}\cdot\cos\vartheta/{2}={2\pi}(l+\frac{1}{2})/{|\bm{q}|},\l=(1,2,3,...).
\end{eqnarray}
%
That's interesting because we can deduce the internuleus distance by
estimating the momentum transfer through the conservation
relationship (14). It's obvious that $\zeta(\bm{q},\bm{R})$ is at
its peak at the resonances regardless of the internuclear distance.
This can be explained by considering that the momentum transfer onto
the nucleus is almost zero when the resonance condition is satisfied
(i.e., the intermediate electron becomes real). That's to say the
resonance peak of the spectrum carry little information about the
internuclear distance or orientation, for which we will not pay much
attention to the phenomenon of resonance.

\begin{figure}
\includegraphics[scale=0.5]{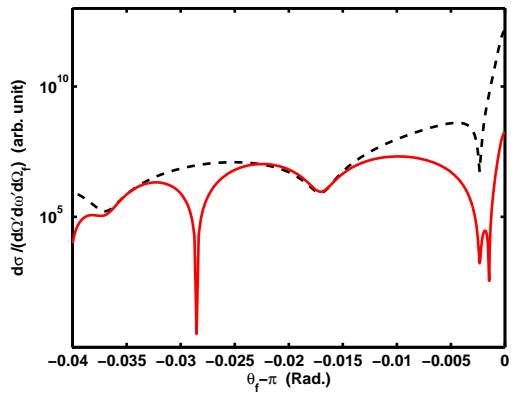}
\caption{\label{fig:differential}The differential cross section as a
function of the electron emission angle $\theta_f$ for the
fundamental harmonic for absorbing 1 photon in the whole process
(n=1). The parameter of the electron and laser is the same with Fig
~\ref{fig:cross-section}. The internucleus distance is 1nm for the
full line, 0.1nm for the dash line.}
\end{figure}

\begin{figure}
\includegraphics[scale=0.5]{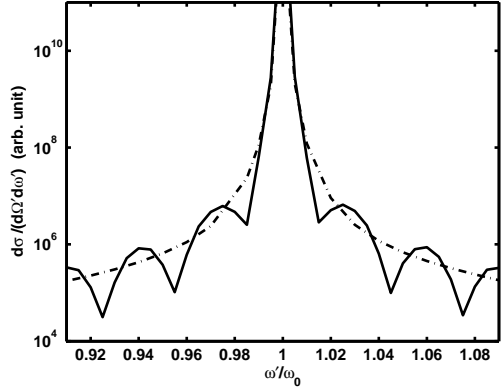}
\caption{\label{fig:cross-section1}The cross section for the
fundamental harmonic at $\theta^\prime=1^\circ$. The parameter of
the electron and laser is the same with Fig3. The internucleus
distance is 1nm and the orientation is ($\vartheta=0$)for the full
line and ($\vartheta={\pi}/{2}$) for the dash line.}
\end{figure}

The dependence of the differential cross section in (18) on the
electron emission angle $\theta_f$ at frequency
$\omega^\prime=0.955\omega_0$ is plotted in Fig
~\ref{fig:differential}. It located close to one of the minima in
the spectrum. We observe a clear suppression of the emission at
small angle around the direction of the ingoing electron for
$R_0=1nm$, which result in the minimum of the spectrum.
%
As can be expected from (19), if we increase the angle between the
orientation of the two nuclei and the direction of the laser
propagation, the two-centre interference will be less effective.
Finally, we even could not find a pronounced minimum in the spectrum
when the internucleus orientation is perpendicular to the direction
of the laser propagation ($\vartheta=\frac{\pi}{2}$). In order to
corroborate this idea, we plot the full cross section for
$\vartheta={\pi}/{2}$ in comparison with that of $\vartheta=0$  for
the same internucleus distance in Fig ~\ref{fig:cross-section1}. The
explanation for this phenomenon is that the momentum transfer from
the Coulomb field q is mainly in the backscattering direction
according to the conservation relationship (14), thus
$\bm{q}\cdot\bm{R}\approx0$. We can conclude that the more
projection of the internucleus distance onto the laser propagating
direction, the more oscillation occurs for the parameter
$\zeta(\bm{q},\bm{R})$ , which leads to the appearance of the minima
in spectrum.
%
The initial velocity of the ingoing electron also has a large effect
on this two-centre interference phenomenon. That's because it will
influence the momentum transfer from the Coulomb field to the
electron. Here we still set ($\vartheta=0$) to maximize the
two-centre interference. To have a clear idea of the relation
between the initial velocity $v_i$ and the momentum transfer from
the Coulomb field on the internucleus orientation $q^4$, we shall
calculate the derivative ${dq^4}/{dv_i}$ for fundamental harmonics.
From the conservation relationship, we have (here we set
$\hbar=m=c=1$ ):

\begin{eqnarray}
\frac{dq^4}{dv_i}=\big[\frac{1-\frac{1+Q^2/2}{(\pi_i^0)^2}}{1-\frac{1+Q^2/2}{(\pi_i^0+\Delta\omega)^2}}\big]^{1/2}\cos(\theta_f-\pi)-1,
\end{eqnarray}

\begin{figure}
\includegraphics[scale=0.5]{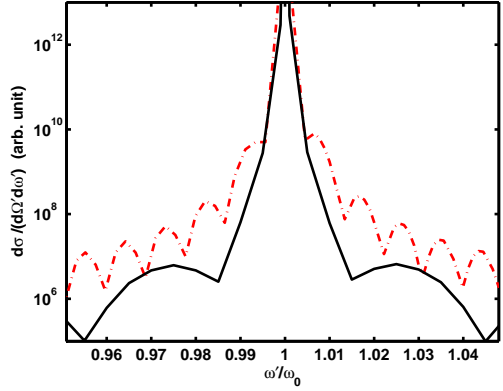}
\caption{\label{fig:cross-section2}The cross section for the
fundamental harmonic at $\theta^\prime=1^\circ$. The parameter of
the electron and laser is the same with Fig 3 except that the
initial energy of the electron reduces to 3.5MeV for the dotted-dash
line and 5MeV for the full line. The internucleus distance is 1nm.}
\end{figure}
%

Here $\Delta\omega=\omega^\prime-\omega_0$. For $\theta_f\approx
\pi$, we could learn there will be more momentum transfer for
smaller initial velocity based on (19). Remembering the interference
is in connection with the term $\zeta(\bm{q},\bm{R})\sim
\cos^2(\bm{q}\cdot\bm{R})$, so we expect there will be more minima
on the spectrum for "slow" electron but still satisfying the Born
approximation, as can be seen from Fig ~\ref{fig:cross-section2}.
Here we compare the spectrum for initial electron energy
$E_i=3.5MeV$ with that of $E_i=5MeV$. The locations of the minima on
the spectrum are different and the interval is smaller for
$E_i=3.5MeV$, which confirms our opinion.For the same reason, we
expect this will also happen with the increasing laser intensity
since the electron is more decelerated by the light pressure of a
counterpropagating laser.

It has to be mentioned that the parameter ($Q=17.8$ with a
wavelength of 0.2nm) we choose in the paper will correspond to a an
X-ray laser with intensity up to $I=10^{28}W/cm^2$, which, according
to an optimistic view\cite{19}, could be reached with future
upgrades of the FLASH facility in Hamburg. On the other hand,
considering a neodymium laser with a frequency of 1.17eV and
$Q=17.8$ (corresponding to an intensity $I=10^{28}W/cm^2$), the
clear interference effect of laser-assisted bremsstrahlung emission
also could be found when $\bm{q}\cdot\bm{R}\sim1$ is satisfied.
That's to say, the corresponding internucleus distance has to be on
an order of micrometer ($\bm{R}\sim10^{-6}m$) according to our
calculation. Moreover, it is true that the interference modulations
may be also found in the Bethe-Heitler cross section generated by
the electron scattering in multi-center potentials in absence of a
laser field. Considering the appearance of the resonances, we may
expect a great difference between the interference modulations of
the laser-free spectrum from those of the laser-assisted spectrum we
found in this paper(i.e, there will not be minima located
symmetrically on each side of the resonance in the the laser-free
spectrum). The further discussion of the detail about the
differences between the two spectrum is beyond the topic of this
paper. Furthermore, it's possible to modulate the intensity and the
interference diagram of the electron radiation spectrum by
controlling the laser field in an actual experiment, for which we
think the laser field is helpful in observing a clear interference
effect in the spectrum.
\\


\section*{\romannumeral4. Conclusion}
In this paper, we have investigated the scattering of an electron by
the screened Coulomb field of two fixed nuclei in a highly intense
laser field and then emit a bremsstrahlung photon. As a result, we
found that the spectrum may exhibit minima away from the resonant
frequency. This may be explained by the interference between
contributions from two fixed nuclei. It was shown that the positions
of the interference minima are characteristic of both the
internuclear distance and orientation for given laser and electron.
On the other hand, the laser intensity and wavelengthrference, the
initial electron energy is also responsible for the observed minima.
It is shown that the interference effect is remarkable for slow
electrons counter-propagating with the laser field.  That is due to the
large momentum transfer from the Coulomb field. This interference
effect is very general in highly intense laser field in which the
drift motion of the electron can not be neglected. Choosing proper
laser wavelength, one can obtain information about the molecule
structure by detecting the corresponding photon spectrum.

Finally, we must point out that the idea discussed in this paper about the two-center potential can be generalized to more complex
potentials to find its important practical applications. For example, the existence of multiply-charged negative ions in intense high-frequency laser fields has been studied theoretically for a long time \cite{20}. Recently, by including relativistic corrections, the ions have been found to be able to bind more electrons \cite{21}. How to detect these exotic ions existing only in intense laser fields has posed a great challenge to present experimentalists. Our next work will focus upon analyzing the characteristics of the radiation spectrum when the free electrons are injected upon the negative ions inside the laser fields, for these ions have a very special dressed potential structure just like a many-atom molecule.

\acknowledgements This work is supported by the National Natural Science
Foundation of China under Grant Nos. 10974056 and 11274117. One of the authors, Wenjun Zhu, thanks the support by the Science and Technology Foundation of National Key Laboratory of Shock Wave and Detonation Physics (Grant No. 077110).\\

\appendix

\section{Derivation of Eq.(11)}

Considering the expression (2),(3),(4) and (9),  the transition
amplitude of an electron scattering (8) reads as:
\begin{eqnarray}
S_{fi}=-\frac{e^2c}{\hbar^2
c^2}\sqrt{\frac{m^2c^2}{\Pi_i^0\Pi_f^0V}}\sqrt{{2\pi\hbar}/{\omega^\prime}}T_{fi},\nonumber
\end{eqnarray}

Here
\begin{eqnarray}
&&T_{fi}=T_{fi}^{(1)}+T_{fi}^{(2)}\nonumber\\
&&T_{fi}^{(1)}={\int dx^4dy^4\bar{u_{r_f}(p_f)}[\bar{\zeta}_{p_f}(x)
\rlap{$\slash$}\epsilon_c \zeta_{p}(x)]iS(x-y)[\bar{\zeta}_{p}(y)
\rlap{$\slash$}{A_Y}(y) \zeta_{p_i}(y)]u_{r_i}(p_i) e^{i{k^\prime}x}}\nonumber\\
&&iS(x-y)=-\frac{1}{2\pi
i}{\int\frac{d^4p}{(2\pi\hbar)^3}\frac{1}{{\rlap{$\slash$}p-mc}}}
\end{eqnarray}

{$T_{fi}^{(2)}$} can be obtained from {$T_{fi}^{1}$} by
interchanging: {$x\to y,
\rlap{$\slash$}\epsilon_c\to\rlap{$\slash$}A_Y$}

With the definition of {$\zeta_p(x)$} in (3), it follows the
relation:

\begin{eqnarray}
\bar{\zeta}_{p_f}(x) \rlap{$\slash$}\epsilon_c
\zeta_{p}(x)&=&\sum_{s_1}
{M_{s_1}(\rlap{$\slash$}\epsilon_c,\eta^1_{\Pi,\Pi_f},\eta^2_{\Pi,\Pi_f})e^{i(\Pi_f-\Pi+s_1\hbar
k)x/\hbar}}\nonumber \\
\bar{\zeta}_{p}(y)\rlap{$\slash$}A_Y(y)\zeta_{p_i}(y)&=&\sum_{s_2}
{\bar{M_{s_2}}(\rlap{$\slash$}A_Y(y),\eta^1_{\Pi,\Pi_i},\eta^2_{\Pi,\Pi_i})e^{i(\Pi_i-\Pi+s_2\hbar
k)y/\hbar}}
\end{eqnarray}

During the calculation, the following expression will be useful:
\begin{eqnarray}
exp\big(i\alpha\sin(kx-\varphi)-i\beta\sin{2kx}\big)=\sum_s {G^0_s(\alpha,\beta,\varphi)e^{iskx}}\nonumber \\
\cos(kx)exp\big(i\alpha\sin(kx-\varphi)-i\beta\sin2kx\big)=\sum_s {G^1_s(\alpha,\beta,\varphi)e^{iskx}}\nonumber \\
\sin(kx)exp\big(i\alpha\sin(kx-\varphi)-i\beta\sin2kx\big)=\sum_s {G^2_s(\alpha,\beta,\varphi)e^{iskx}}\nonumber \\
\sin(2kx)exp\big(i\alpha\sin(kx-\varphi)-i\beta\sin2kx\big)=\sum_s {G^3_s(\alpha,\beta,\varphi)e^{iskx}}\nonumber \\
\end{eqnarray}

All integrations can be taken in the expression of {$T_{fi}^{(1)}$},
leaving the energy-conserving delta function:
{$\delta(\Pi_f-\Pi+s_1\hbar k+\hbar k^\prime)$} and
{$\delta(\Pi_i-\Pi+s_2\hbar k+\hbar q)$}, which leads to the
energy-momentum conserving relation(14). Finally, the expression for
{$T_{fi}^{(1)}$} reads:

\begin{eqnarray}
T_{fi}^{(1)}=&&(\hbar^2)\sum_{s,n}
A_Y^0(q)\bar{u_{r_f}(p_f)}M_{-n-s}(\rlap{$\slash$}\epsilon_c,\eta^1_{\Pi,\Pi_f},\eta^2_{\Pi,\Pi_f})
\frac{i}{\rlap{$\slash$}p-mc}\nonumber\\
&&\times\bar{M_{-s}}(\gamma^0,\eta^1_{\Pi,\Pi_i},\eta^2_{\Pi,\Pi_i})u_{r_i}(p_i)
\delta(\Pi_f^0-\Pi^0-n\hbar k^0-\hbar k^{\prime 0}).
\end{eqnarray}

The expression of {$T_{fi}^{(2)}$} is similar to that of
{$T_{fi}^{(2)}$} by substitutions as follows:
\begin{eqnarray}
{\Pi\to\Pi^\prime},{\rlap{$\slash$}\epsilon_c\to\gamma^0}\nonumber.
\end{eqnarray}

Taking the square of the transition amplitude {$S_{fi}$}, we will
finally have the expression (11).


\end{document}